\documentclass[12pt,preprint]{aastex}

\slugcomment{}

\shorttitle{NLTE \ion{Ba}{0} and \ion{Sr}{0}}
\shortauthors{Short and Hauschildt}

\begin{document}


\title{NLTE Strontium and Barium in metal poor red giant stars}


\author{C.I. Short}
\affil{Department of Astronomy \& Physics and Institute for Computational Astrophysics, Saint Mary's University,
    Halifax, NS, Canada, B3H 3C3}
\email{ishort@ap.smu.ca}

\author{P.H. Hauschildt}
\affil{Hamburger Sternwarte, Gojenbergsweg 112, 21029 Hamburg, Germany}
\email{phauschildt@hs.uni-hamburg.de}


\begin{abstract}

  We present atmospheric models of red giant stars of various 
metallicities, including extremely metal poor (XMP, 
$[{{\rm Fe}\over{\rm H}}]<-3.5$) 
models, with many chemical species, including, significantly, 
the first two ionization stages of Strontium (\ion{Sr}{0}) and Barium 
(\ion{Ba}{0}), treated in Non-Local Thermodynamic Equilibrium (NLTE)
with various degrees of realism.  
We conclude that 1) for all lines that are useful \ion{Sr}{0} and \ion{Ba}{0} abundance diagnostics
the magnitude and sense of the computed NLTE effect on the predicted line strength 
is metallicity dependent, 2) the indirect NLTE effect of overlap between 
\ion{Ba}{0} and \ion{Sr}{0} transitions and transitions of other species that are 
also treated in NLTE non-negligibly enhances NLTE abundance corrections for some lines,
3) the indirect NLTE effect of NLTE opacity of other species on the equilibrium structure of 
the atmospheric model is {\it not} significant, 
4) the computed NLTE line strengths differ negligibly if
collisional $b-b$ and $b-f$ rates are an order of magnitude smaller or larger than those 
calculated 
with standard analytic formulae, and 5) the effect of NLTE upon the resonance line of 
\ion{Ba}{2} at 4554.03 \AA~
is independent of whether that line is treated with hyperfine splitting.   
As a result, the derivation of abundances of \ion{Ba}{0} and \ion{Sr}{0} for
metal-poor red giant stars with LTE modeling that are in the literature 
should be treated with caution.

\end{abstract}


\section{Introduction}

Generally, the time development of the neutron capture ($n$-capture) element ($Z>30$)
abundances
over the galaxy's history encodes information about the astrophysical sites of 
the various $n$-capture processes, such as the ``classical'' rapid process ($r$-process), 
and the weak and main slow processes ($s$-processes), and their relation
to the sites of light metal ($Z<30$) nucleosynthesis (see \citet{cowan_t04} for a
general review).  High quality spectroscopy of metal-poor and extremely metal-poor
(XMP) red giants in the halo component of the Milky Way galaxy has allowed the determination
of $n$-capture element abundances at very early times in the formation of the galaxy
(see \citet{james_fbcgs04} (hereafter: JFBCGS04), for a recent example).  
In particular, the abundances of \ion{Sr}{0} and \ion{Ba}{0} are significant because,
if their formation is understood correctly, they may
indicate the relative rate of nucleosynthesis 
of light and heavy $n$-capture elements, and, by extension, the relative importance of
the weak and main $s$-processes in the chemical evolution of the galaxy (see JFBCGS04
and references therein). 

\paragraph{}

Almost all large scale abundance surveys of old red giants to date have been carried out 
with atmospheric models and synthetic spectra computed under the assumption of
Local Thermodynamic Equilibrium (LTE), either with MARCS models \citep{edvardsson_aglnt93}
or ATLAS9 models \citep{kurucz94}.  Generally, previous investigation has indicated that 
the equilibrium state,
and, thus, the line opacity, of many elements may deviate from LTE, particularly 
in giant atmospheres where gas densities, and, hence, thermalizing collision rates, 
are low (see \citet{short_h03} for example).  Furthermore, it has been found that 
the extent of NLTE effects depends on the overall metallicity, $[{{\rm A}\over{\rm H}}]$
of the atmosphere.
\citet{mashonkina_gb99}, (hereafter: MGB99), carried out Non-LTE (NLTE) modeling of the 
\ion{Ba}{2} 
lines for main sequence
and turn-off stars using a complete linearization method to solve the NLTE statistical
equilibrium (SE) rate equations given an LTE atmospheric model structure.   The study of
MGB99 is an important pioneering calculation, and concluded that NLTE
effects affected the inferred abundance of \ion{Ba}{0} by as much as $0.2$ dex for 
stars of $[{{\rm A}\over{\rm H}}]<-2$.  However, there are three things about the study
of MGB99 that we draw attention to: 1) it is not self-consistent in that the
model structure adopted does not take into account NLTE effects on the many other
species, such as Fe, whose opacities determines the radiative equilibrium structure,
2) it is not self-consistent in that other species that may have overlapping transitions
with \ion{Ba}{0} are treated in LTE,  
3) because of the inefficiency of the complete linearization method the \ion{Ba}{2}
atomic model is incomplete, having 35 levels and (judging by the Grotrian diagram in their
Fig. 1) 28 bound-bound ($b-b$) transitions. 
It is not clear that these limitations
affect the results significantly without performing numerical experiments with the 
same models.  However, as described in Section \ref{sec_model}, 
our NLTE modeling should be more realistic on all three of these points. 

\paragraph{}

The atmospheric and spectrum modeling with the {\tt PHOENIX} code
for the mildly metal-poor red giant Arcturus
($\alpha$ Boo, HD124897, spectral type K2\,{\sc III}) presented by
\citet{short_h03} included many
thousands of lines of many species in NLTE, but did not include \ion{Sr}{0} or \ion{Ba}{0}.
We have recently added to the NLTE atmospheric and spectrum synthesis code {\tt PHOENIX}
the ability to compute both of these species in NLTE.  The purpose of the present
investigation is to determine the effect on the computed \ion{Sr}{0} and \ion{Ba}{0} 
lines, and, hence, on the inferred abundances, of three types of NLTE consideration: 
1) direct NLTE effects on the \ion{Sr}{0} and \ion{Ba}{0} equilibrium and line formation, 2) indirect
NLTE effects on the atmospheric structure from NLTE effects on the opacity of many
other chemical species, and 3) the indirect NLTE effect on the computed \ion{Sr}{0} or \ion{Ba}{0}
state by overlap of \ion{Sr}{0} and \ion{Ba}{0} transitions with transitions of other 
species also in NLTE.  Moreover, we investigate how the NLTE effects vary with
overall value of $[{{\rm A}\over{\rm H}}]$, including XMP values.  Here, the investigation is carried out 
for the most readily observable metal poor stars in the galactic halo: red giants. 
In Section \ref{sec_model} we describe the model calculations,
in Section \ref{sec_reslt} we present our results, and we re-iterate our main 
conclusions in Section \ref{sec_con}.

\section{Modeling \label{sec_model}}

\subsection{Stellar parameters}

To represent a typical XMP red giant of the type that has been observed in the galactic 
halo we adopt the following stellar parameters: a $T_{\rm eff}$ value of 4800 K, a $\log g$ value 
of $1.5$, a value for the microturbulent velocity dispersion, $\xi_{\rm T}$, of 2.0 km s$^{-1}$,
and a convective mixing length parameter of 2 pressure scale heights.  The models are 
computed with spherical geometry with a radius at the $\tau_{\rm 12000}$ surface of
$1.6\times 10^{12}$ cm (~25 R$_{\odot}$), where $\tau_{\rm 12000}$ is the optical depth due
to {\it continuous} opacity at a standard reference wavelength of 12000 \AA.  These parameters were
chosen to be approximately representative of two XMP red giants recently described by \citet{cayrel_ds04}:
CD\,-38:245 and CS\,22949-037.  

\paragraph{Abundances: }
A set of models of these parameters was generated with a scaled
solar abundance distribution spanning a 
range in $[{{\rm A}\over{\rm H}}]$ value from -1 (mildly metal poor) to -5 (XMP), with an interval
of -1.  We adopt the solar abundances of \citet{grev_ns92}. 
In all cases the values of $[{{\rm Sr}\over{\rm H}}]$ and $[{{\rm Ba}\over{\rm H}}]$ 
equal the value of $[{{\rm A}\over{\rm H}}]$.   The use of a scaled solar abundance distribution
is known to be unrealistic for Pop II, and especially extreme Pop II, stars.  However, 
there is a range of abundance distributions that has been found for XMP stars, so no one choice
would represent all of them.  Furthermore, the purpose of our study is to investigate
{\it differential} effects of $[{{\rm A}\over{\rm H}}]$ value on the NLTE spectrum formation of 
\ion{Sr}{0} and \ion{Ba}{0}, and, to begin with, we suppose that the abundance distribution will
be unimportant to first order for such differential effects.     

\subsection{{\tt PHOENIX}}

{\tt PHOENIX} makes use of a fast and accurate Operator Splitting/Accelerated Lambda Iteration
(OS/ALI) scheme to solve self-consistently the radiative
transfer equation and the NLTE statistical equilibrium (SE) rate equations for many species and overlapping transitions
\citep{hauschildt_b99}.  In our fully self-consistent models (NLTE$_{\rm Full}$ models,
described below) we
include a total of 39 species in the set of multi-level NLTE SE equations, including
the lowest one to three ionization stages of most light metals up to and including
the Fe group, as well as the first two ionization stages of \ion{Sr}{0} and \ion{Ba}{0}.  
We have only included ionization stages that are significantly populated at some depth in
the atmosphere of a late-type star, as computed in radiative-convective equilibrium ({\it ie.}
no chromosphere or transition region).  Therefore, no ionization stages higher than \ion{ }{3}
are treated in NLTE for any species. 
 \citet{short_h05} contains details of the number of ionization stages, $E$ levels, and
$b-b$ and bound-free ($b-f$) transitions included in the NLTE {\tt PHOENIX} SE calculation for
late-type stars, as well as 
the sources of atomic data and the formulae for various atomic processes.  

\subsection{Degrees of realism}

Table \ref{t1} shows the four degrees of realism in our
modeling:

\paragraph{}
1) NLTE$_{\rm Full}$: the most self-consistent NLTE models with 35 other species 
in addition to \ion{Sr}{0} and \ion{Ba}{0} \ion{ }{1} and \ion{ }{2} included in NLTE in the calculation of {\it both}
the atmospheric $T$ and $N_{\rm e}$ structure {\it and} the equilibrium state of \ion{Sr}{0} and \ion{Ba}{0}.
Included in NLTE SE are species such as \ion{Fe}{1} and \ion{ }{2} that are 
known to have an important
effect on the computed radiative equilibrium $T$ structure.
These models account for the {\it indirect} effects of {\it both} the NLTE opacity of many species on
the atmospheric $T$ and $N_{\rm e}$ structure, {\it and} multi-level NLTE effects caused by transitions of other
species in NLTE that overlap with \ion{Sr}{0} and \ion{Ba}{0} transitions, as well as
the {\it direct} effects of NLTE on \ion{Sr}{0} and \ion{Ba}{0} themselves.

\paragraph{}
2) NLTE$_{\rm Sr+Ba}$: {\it only} \ion{Sr}{0} and \ion{Ba}{0} treated in NLTE SE with the 
NLTE$_{\rm Full}$ $T$ structure of 1).  These models account for the effects of NLTE opacity 
of other species on the atmospheric structure, and for NLTE effects on \ion{Sr}{0} and \ion{Ba}{0},
but {\it not} for the effects of NLTE on transitions of other species that overlap those
of \ion{Sr}{0} and \ion{Ba}{0}.

\paragraph{}
3) NLTE$_{\rm LTE}$: deliberately inconsistent models in which the state of \ion{Sr}{0} and \ion{Ba}{0}
is computed in LTE ({\it ie.} {\it no} species in the NLTE SE), {\it but} with the NLTE$_{\rm Full}$
atmospheric structure of 1).  The purpose of these models is to distinguish between effects
on the \ion{Sr}{0} and \ion{Ba}{0} lines that are caused by {\it direct} NLTE effects on the \ion{Sr}{0} and \ion{Ba}{0}
equilibrium and those that are caused by the NLTE atmospheric structure.

\paragraph{}
4) LTE: fully LTE models with the atmospheric $T$ and $N_{\rm e}$ structure and the state
of all species, including \ion{Sr}{0} and \ion{Ba}{0}, computed in LTE.  These are the simplest
models and are equivalent to the MARCS and ATLAS9 models that are largely used in abundance
analyses.

\subsection{\ion{Sr}{0} and \ion{Ba}{0} \ion{ }{1} and \ion{ }{2}}

Table \ref{t2} contains details of the numbers of $E$ levels and $b-b$ transitions of
\ion{Sr}{0} and \ion{Ba}{0} \ion{}{1} and \ion{}{2} included in the NLTE SE
rate equations.  All levels not designated as "Sum" are fine structure ($J$) 
levels, and were included separately in the NLTE SE rate equations.
Energy level values, $\log gf$ values, and damping parameters have been taken from 
\citet{kurucz94}.  Because the purpose of our study is to investigate the 
differential effect of modeling treatments rather than to fit observed spectra,
we have not performed a quality control investigation or fine-tuned the atomic data,
except for the collisional cross-section perturbation analysis that follows. 
As a point of information, we note that many of the fine structure level designations
for these species in 
the atomic data base of \citet{kurucz94} are erroneous.  Among the the levels 
included in our atomic models, all levels of \ion{Ba}{0} 
and select levels of \ion{Sr}{0} were fixed to agree with the designations given in the National   
Institute of Standards and Technology (NIST) Atomic Spectra Database (for \ion{Ba}{0}) or
 Handbook of Basic Atomic Spectroscopic Data (for \ion{Sr}{0}).  Unfortunately, we were
unable to find data for many of the \ion{Sr}{0} levels in any of the readily accessible 
databases, so were unable to check the designations for all the levels.  
As with all the species treated in NLTE, only levels of \ion{Sr}{0} and \ion{Ba}{0} connected by
transitions of $\log gf$ value
greater than -3 (designated primary transitions) are included directly in the SE rate 
equations.
All other transitions of that species (designated secondary transitions) are calculated
with occupation numbers set equal to the Boltzmann distribution value with excitation
temperature equal to the local kinetic temperature, multiplied by the ground state
NLTE departure co-efficient for the next higher ionization stage.

\section{Results and Discussion \label{sec_reslt}}

\subsection{NLTE$_{\rm Full}$ {\it vs} LTE}

\begin{figure}
\plotone{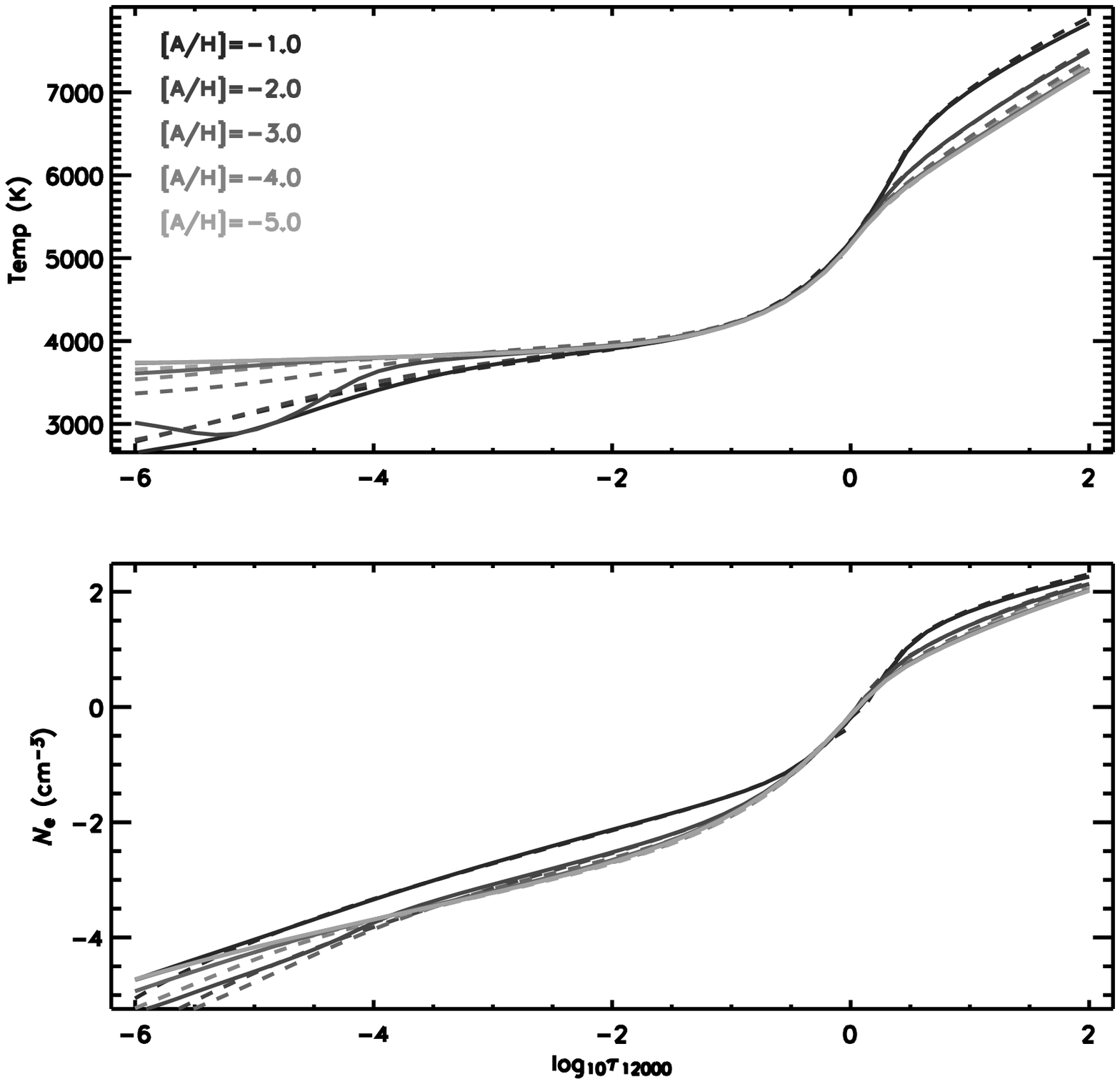}
\caption{Computed $T$ and $N_{\rm e}$ structures of the atmospheric models as
a function of the continuum optical depth at 12000 \AA.
NLTE$_{\rm Full}$: solid line; LTE: dashed line.
\label{models}}
\end{figure}

Fig. \ref{models} shows the computed atmospheric $T$ and $N_{\rm e}$ structures for
the NLTE$_{\rm Full}$ and LTE models for all values of $[{{\rm A}\over{\rm H}}]$.
The NLTE$_{\rm Full}$ models include much of the \ion{Fe}{0} group (the first two
ionization stages of \ion{Ti}{0}, \ion{Mn}{0}, \ion{Fe}{0}, \ion{Co}{0}, and \ion{Ni}{0})
in NLTE.  These species, especially \ion{Fe}{0}, contribute much of the line opacity
to the blue and UV bands and have a large effect on the computed radiative equilibrium of 
the atmosphere.  Treating them in NLTE generally decreases $T$ near the bottom of 
the atmosphere and increases it near the top of the atmosphere, thereby exhibiting
a reduction in the classical ``back-warming'' effect of line opacity.  
However, the difference in NLTE and 
LTE $T$ structures is negligible throughout much of the line forming region near
$\log\tau_{\rm 12000}\approx 1$.  See \citet{short_h05} and \citet{anderson} for a detailed 
discussion of thermal equilibrium in late-type atmospheres with massive scale NLTE. 
There is a rapid increase in the boundary temperature at the top of the model between
$[{{\rm A}\over{\rm H}}]$ values of -1 and -3, and the model of $[{{\rm A}\over{\rm H}}] = -2$
shows complicated behavior between $\log\tau_{\rm 12000}$ values of -4 and -6.  However,
the $[{{\rm A}\over{\rm H}}] = -2$ model has converged at all depths to a stable equilibrium, and 
its behavior is caused by $[{{\rm A}\over{\rm H}}] = -2$ being a critical value for the
thermal equilibrium in the outer atmosphere.

\paragraph{}

\begin{figure}
\plotone{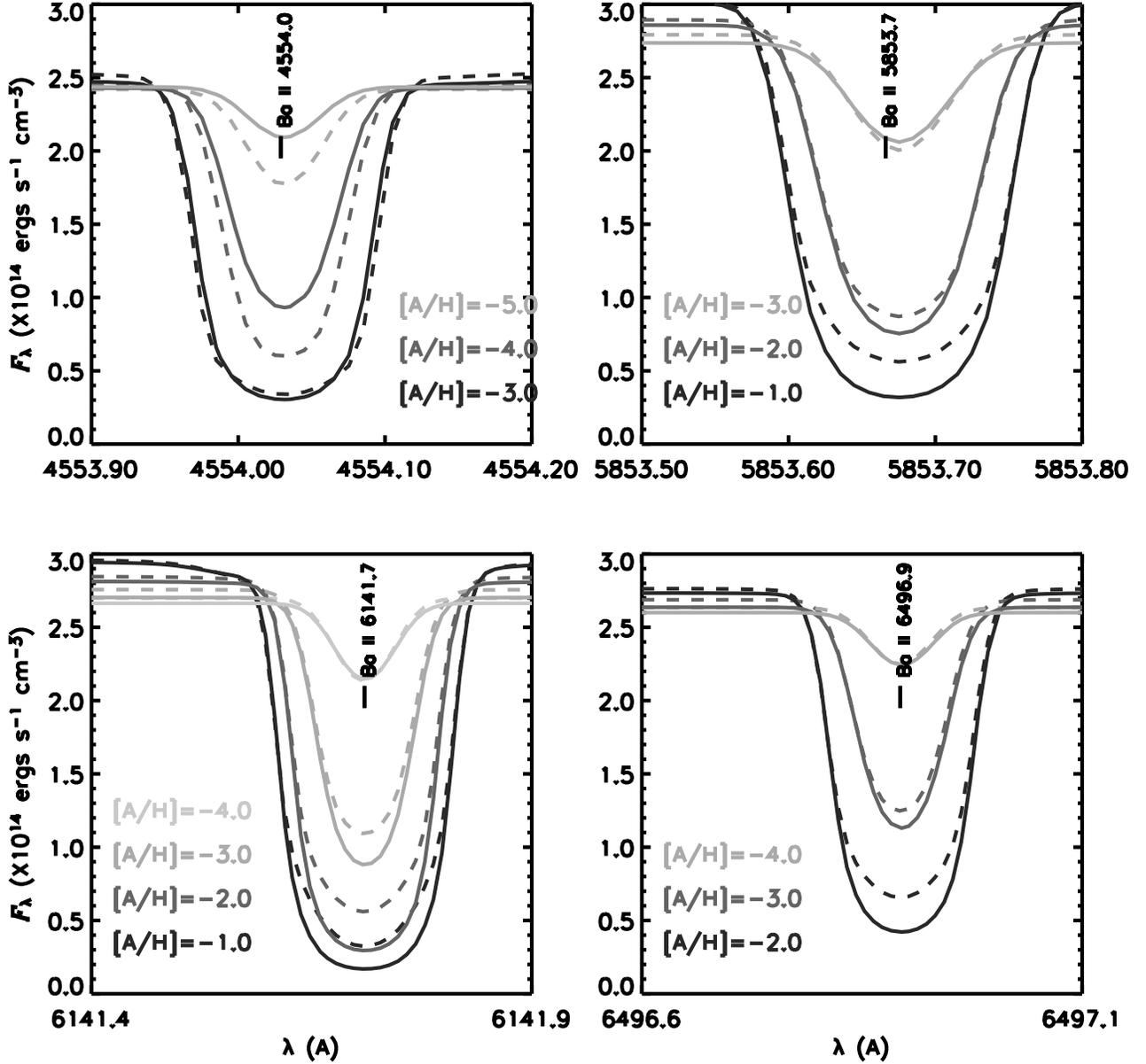}
\caption{Selected computed \ion{Ba}{2} lines for models of $[{{\rm A}\over{\rm H}}]$ values as
indicated on each panel.  Full NLTE model: solid line; 
LTE model: dashed line.  Upper left: $\lambda 4554.03$ (resonance);
upper right: $\lambda 5853.69$; lower left: $\lambda 6141.73$; and lower right $\lambda 6496.91$.  
Note that for each line we only show profiles for models of $[{{\rm A}\over{\rm H}}]$ value
such that the line is of detectable strength, yet unsaturated or only barely saturated.
\label{ba24555}}
\end{figure}

\begin{figure}
\plotone{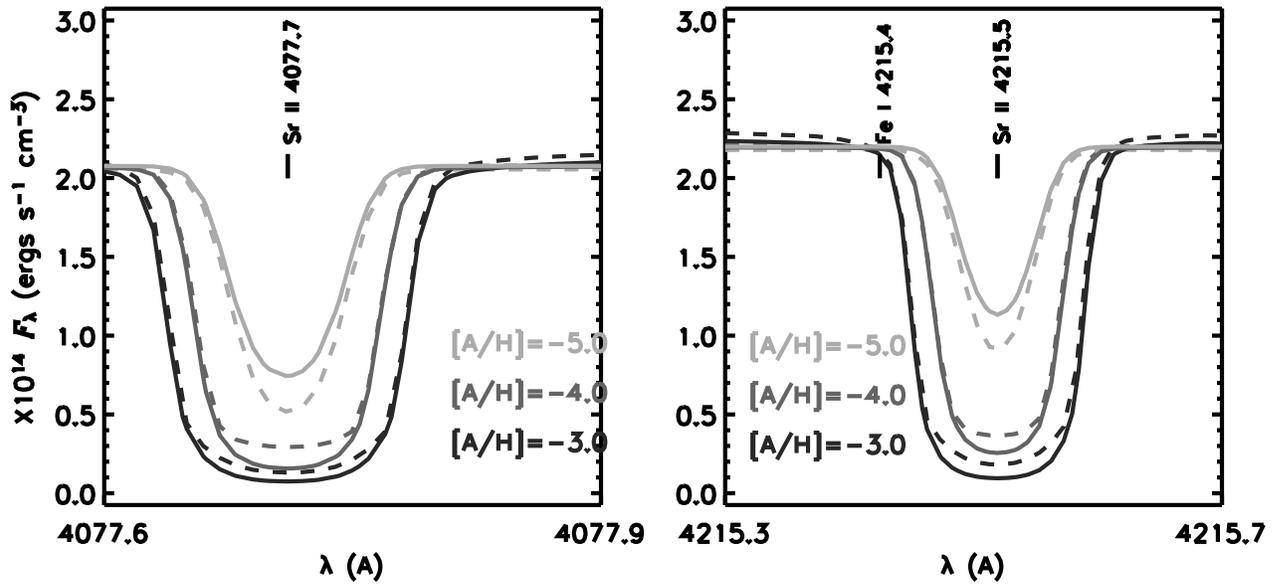}
\caption{As for Fig. \ref{ba24555}, but for \ion{Sr}{2}.  Left panel: $\lambda$ 4077.71;
right panel: $\lambda$ 4215.52. 
\label{sr24217}}
\end{figure}

Figs. \ref{ba24555} and \ref{sr24217} show computed line profiles for the lines used as
abundance diagnostics by JFBCGS04, with all wavelengths in air: \ion{Ba}{2} $\lambda$
4554.03, 5853.69, 6141.73, and 6496.91, and \ion{Sr}{2} $\lambda$ 4077.71 and 4215.52 computed with 
the most and least realistic modeling, NLTE$_{\rm Full}$ and LTE, respectively.  Selected atomic
data from \citet{kurucz94} for these lines are presented in Table \ref{t2_5}, with the designations
of the atomic fine structure levels fixed to agree with the NIST Atomic Spectra Database or the
Handbook of Basic Atomic Spectroscopic Data.  
For each line we show the profiles for those 
models of $[{{\rm A}\over{\rm H}}]$ value 
for which that particular line is of detectable strength, yet on the linear part of the curve of 
growth (COG) ({\it ie.} unsaturated),
or barely saturated, and thus maximally useful as an abundance diagnostic.  These are also
the lines for which NLTE effects, and the realism of their treatment, will make the greatest 
difference.    
Table \ref{t3} shows equivalent widths, $W_\lambda$, for these lines, computed with the various 
degrees of modeling realism.  Again, for each line we only show $W_\lambda$ values for models of 
$[{{\rm A}\over{\rm H}}]$ value such that
the line is detectably strong, yet unsaturated.  

\paragraph{}

For both the \ion{Ba}{0} and \ion{Sr}{0} lines, the effect of NLTE as compared to LTE is to 
increase the predicted strength, $W_\lambda$, of the line for models of higher $[{{\rm A}\over{\rm H}}]$
value and decreases it for models of lower $[{{\rm A}\over{\rm H}}]$ value.  We find that the
critical $[{{\rm A}\over{\rm H}}]$ value varies from line to line throughout almost the
full range of $[{{\rm A}\over{\rm H}}]$ values spanned by our models.  Furthermore, for each line
studied, the NLTE effect is significant at some value of $[{{\rm A}\over{\rm H}}]$.  
For unsaturated lines, NLTE corrections to $[{{\rm Sr}\over{\rm H}}]$ and $[{{\rm Ba}\over{\rm H}}]$  
are proportional to $\log W_\lambda$ in the sense that a model that predicts too large a value of
$\log W_\lambda$ 
because of inadequate physical treatment will cause the inferred value of $[{{\rm Sr}\over{\rm H}}]$
or $[{{\rm Ba}\over{\rm H}}]$ to be proportionally too small, 
and {\it visa versa}.
We note that these results are qualitatively the same as those of MGB99 in their 
study of main sequence and turn-off stars.

\paragraph{}

\begin{figure}
\plotone{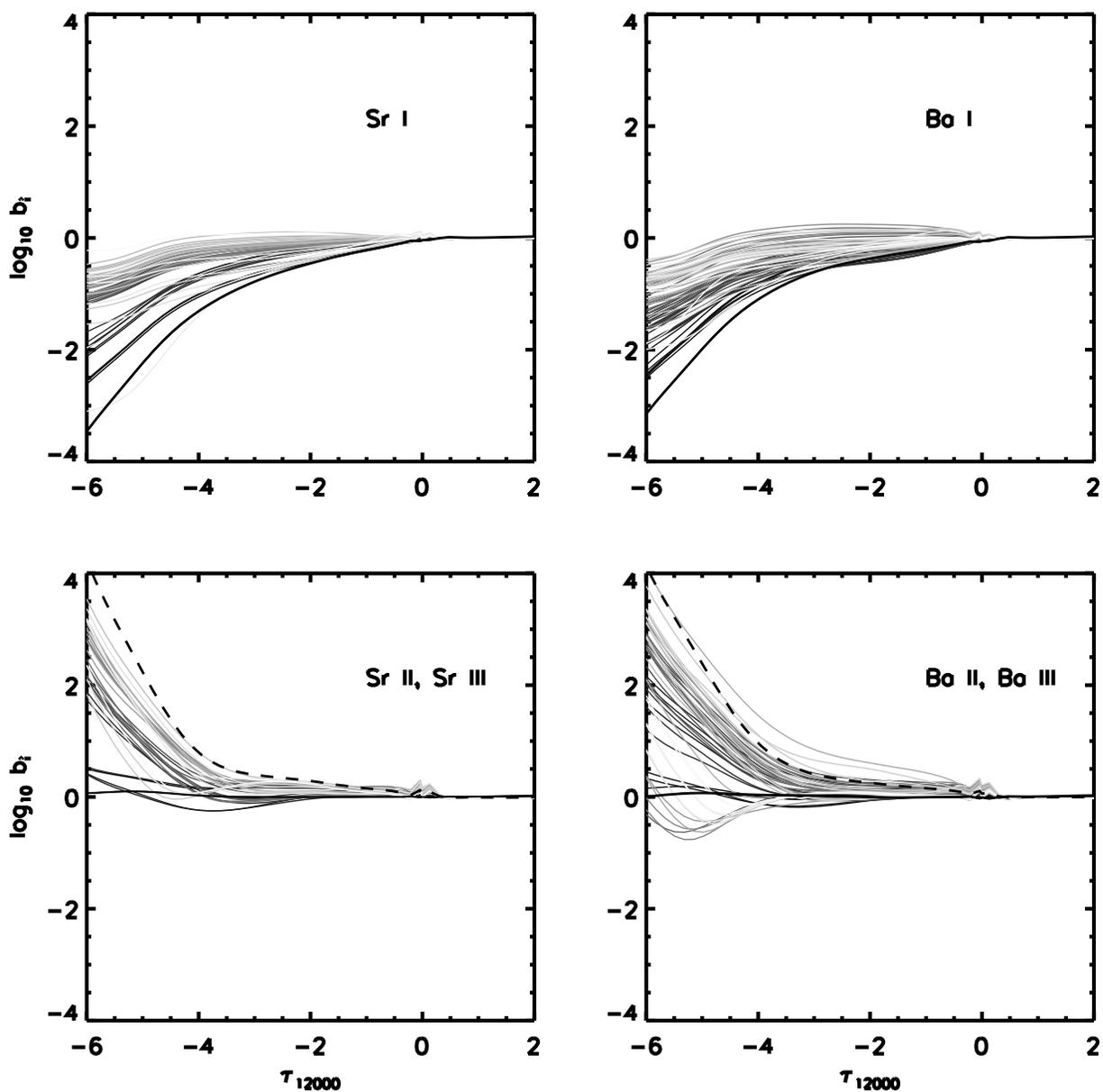}
\caption{Departure coefficients of the NLTE statistical equilibrium 
solution for the model of $[{{\rm A}\over{\rm H}}]=-1$.  
Left panels: \ion{Sr}{0}; right panels: \ion{Ba}{0}; upper panels:
ionization stage \ion{ }{1}; lower panels: ionization stages \ion{ }{2} and \ion{ }{3}.
The ground state coefficient is shown with 
a thick black line.  The lighter
the color of the line the higher the energy, $E$, of the level 
with respect to the ground state.  The dashed line in the lower panels is the ground 
state departure co-efficient of stage \ion{ }{3}.
\label{bi-1}}
\end{figure}

\begin{figure}
\plotone{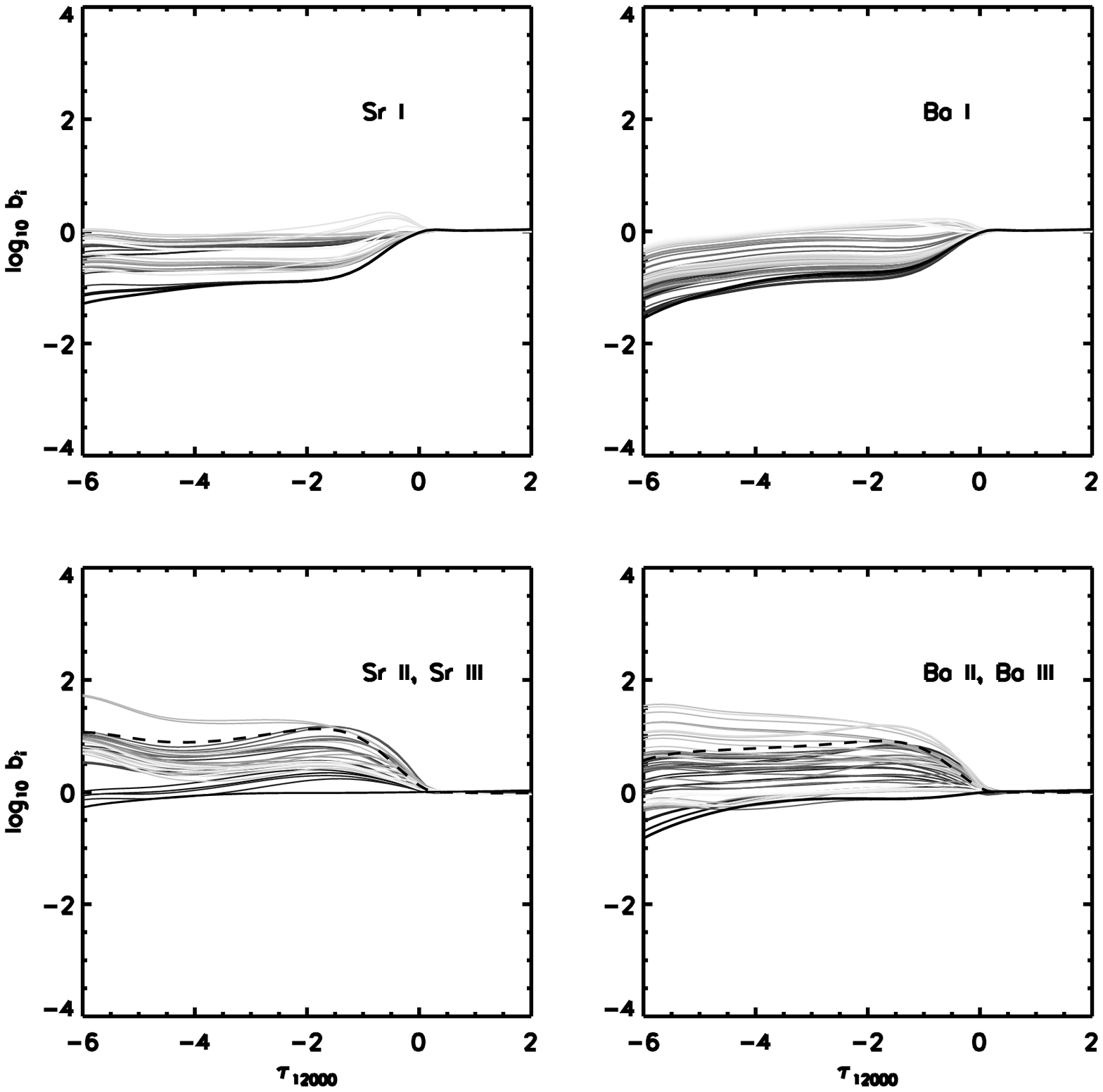}
\caption{As for Fig. \ref{bi-1} except for the model of $[{{\rm A}\over{\rm H}}]=-4$. 
\label{bi-4}}
\end{figure}

Figs. \ref{bi-1} and \ref{bi-4} show for the NLTE$_{\rm Full}$ models of 
$[{{\rm A}\over{\rm H}}]$ value equal to -1 and -4 the logarithm of the NLTE $E$-level departure 
co-efficients, $b_{\rm i}={{n_{\rm i}}\over {n_{\rm i}^*}}$, for each $E$-level, $i$, of 
\ion{Sr}{0} and \ion{Ba}{0} stages \ion{ }{1} and \ion{ }{2}, and the ground state 
of stage \ion{ }{3},
where $n_{\rm i}$ is the actual NLTE occupation number of level $i$ and 
$n_{\rm i}^*$ is the occupation number of the level as computed with LTE
equilibrium relations for all levels and stages with the local value of
$T_{\rm kin}$ and the LTE $N_{\rm e}$.  For the models of $[{{\rm A}\over{\rm H}}]=-1$, 
the $b_{\rm i}$ values for the ground level and the levels of lowest
excitation energy of stage \ion{ }{2} remain close
to unity throughout the atmosphere.  This is expected because stage \ion{ }{2}
is the dominant ionization state and so its total number density,
most of which will be in the ground state, should be negligibly affected by 
departures from the LTE ionization equilibrium.  
However, for the model of $[{{\rm A}\over{\rm H}}]=-4$,
the ground state $b_{\rm i}$ value for \ion{Ba}{2} drops below unity for
$\tau_{\rm 12000}<-4$, and approaches 0.1 at the surface.  
The reason for the difference in the ground state $b_{\rm i}$ behavior
in the models of $[{{\rm A}\over{\rm H}}]=-1$  as compared to those of $-4$
is that in the latter \ion{Ba}{2} is no longer as dominant in the ionization equilibrium,
but is rivaled by \ion{Ba}{3}.
For both $[{{\rm A}\over{\rm H}}]$ values, many levels of higher excitation energy in stage \ion{ }{2}
are overpopulated with respect to LTE ($b_{\rm i} > 1$) in the upper atmosphere
 due to the influence of collisional coupling to the ground level of stage \ion{ }{3}.
The ground level of ionization stage \ion{ }{3} is overpopulated with respect to 
LTE in the outer atmosphere as a result of NLTE overionization of stage \ion{ }{2}
(note that the ground level is the only level of stage \ion{ }{3} included in
our SE equations). 
The ground state of stage \ion{ }{1} is underpopulated with respect
to LTE in the upper atmosphere, again, as a result of NLTE over-ionization. 
In Fig. \ref{ppress} we show the LTE and NLTE partial pressures of the
three ionization stages of \ion{Ba}{0} for models of $[{{\rm A}\over{\rm H}}]$ value 
equal to -1 and -4.  We note that the results for the $b_{\rm i}$ values are consistent 
with the partial pressure results.

\paragraph{}

From a comparison of Figs.
\ref{bi-1} and \ref{bi-4} to Fig. 2 of MGB99
we note that for the models of $[{{\rm A}\over{\rm H}}]=-1$ the results for the ground 
levels of \ion{Ba}{2} and \ion{ }{3} are qualitatively 
similar to theirs for the Sun, where their ground state value of $b_{\rm i}$
for \ion{Ba}{2} remains
close to unity throughout the atmosphere.  
The results for the ground
level of \ion{Ba}{2} for our $[{{\rm A}\over{\rm H}}]=-4$ model are also qualitatively
similar to those of MGB99 for their model of G84-29, which has 
$[{{\rm A}\over{\rm H}}]=-2.6$.  The latter also shows a drop below unity 
in the upper atmosphere.  However, these comparisons are not expected to 
be exact because out models correspond to giant stars, whereas their correspond to 
main sequence stars.


\subsection{NLTE$_{\rm Full}$ {\it vs} NLTE$_{\rm Sr+Ba}$, NLTE$_{\rm LTE}$ and LTE}

\begin{figure}
\plotone{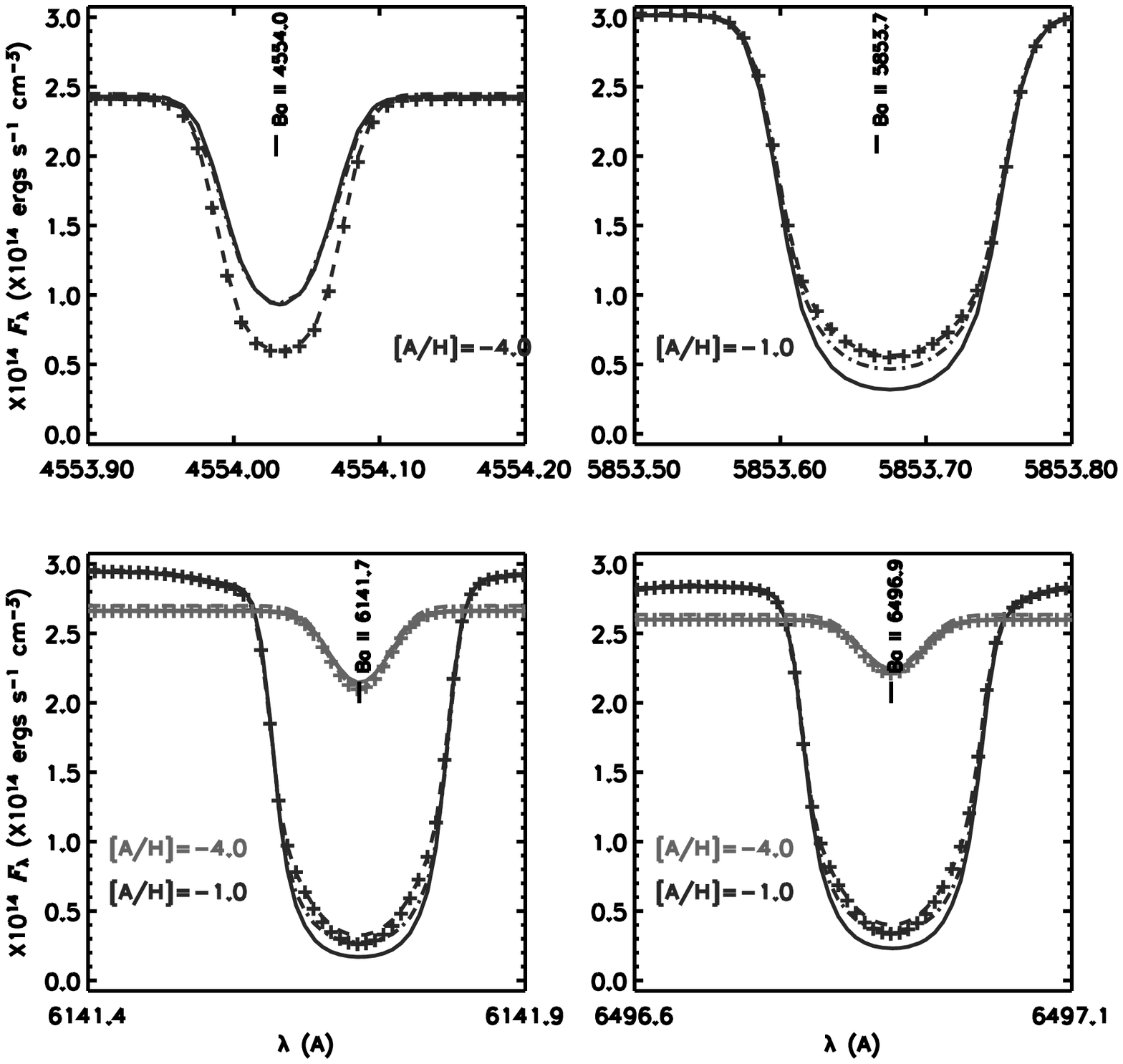}
\caption{As for Fig. \ref{ba24555} with the addition of the lines computed
with the NLTE$_{\rm Sr+Ba}$ (dot-dashed line) and NLTE$_{\rm LTE}$ modeling (crosses). 
\label{ba24555_2}}
\end{figure}

\begin{figure}
\plotone{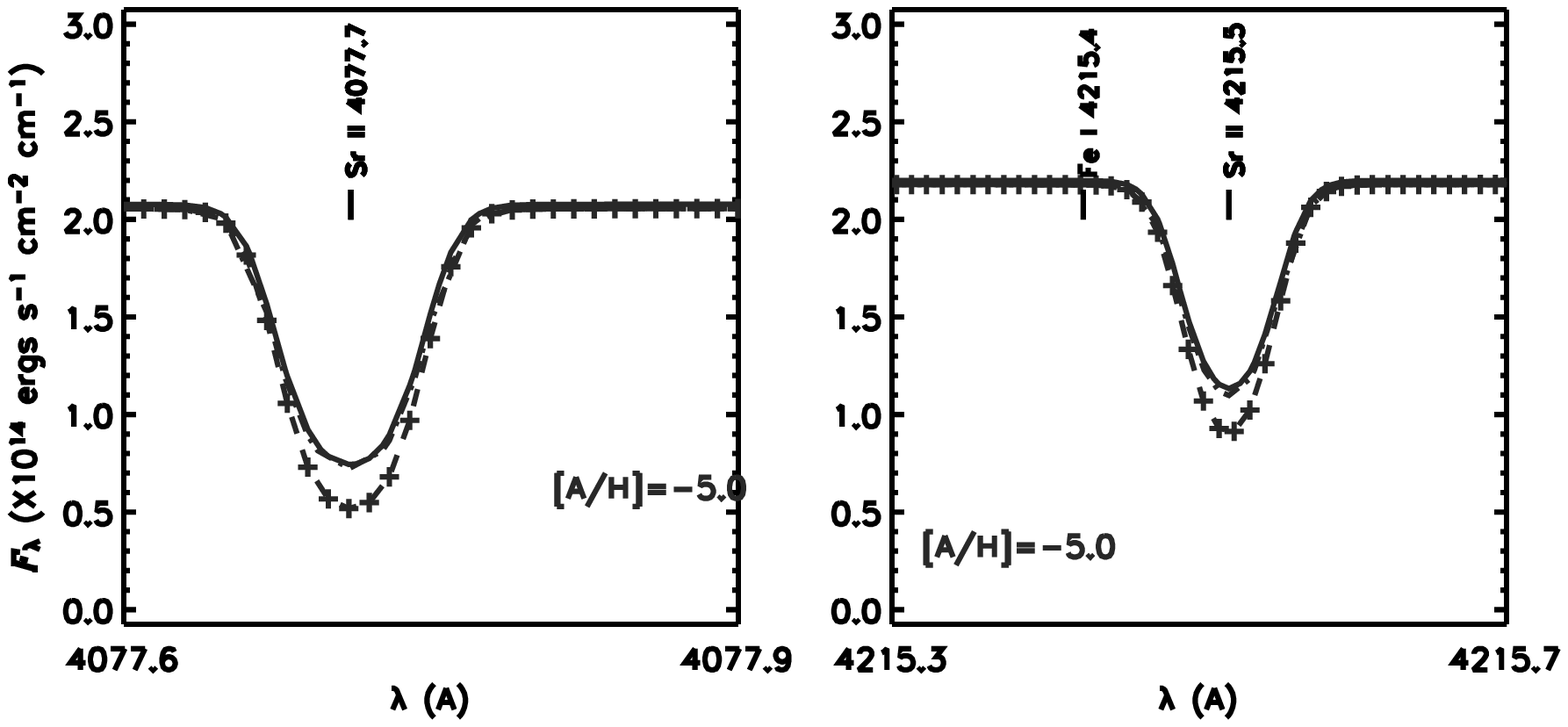}
\caption{As for Fig. \ref{sr24217} with the addition of the lines computed  
with the NLTE$_{\rm Sr+Ba}$ (dot-dashed line) and NLTE$_{\rm LTE}$ modeling (crosses). 
\label{sr24217_2}}
\end{figure}

\begin{figure}
\plotone{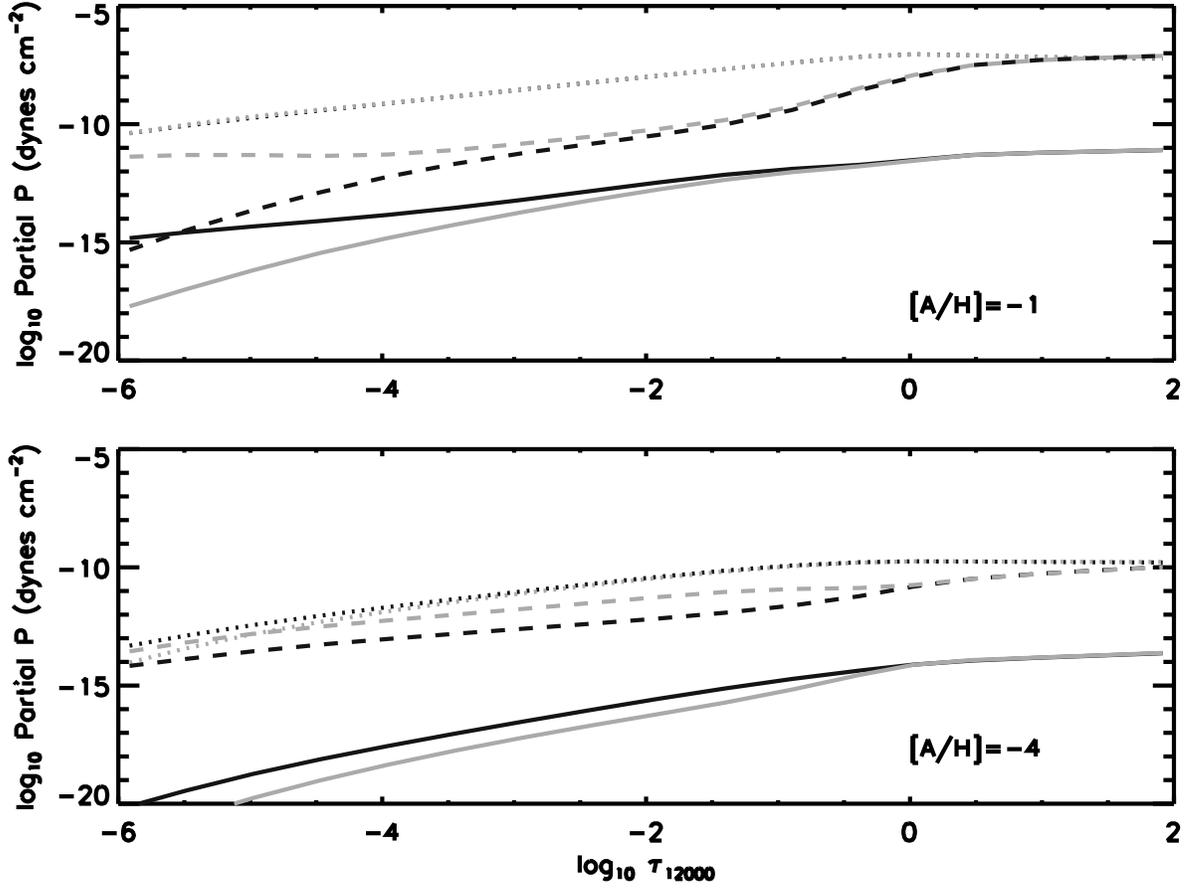}
\caption{Partial pressures of \ion{Ba}{1} (solid line), \ion{Ba}{2} (dotted line), 
and \ion{Ba}{3} (dashed line) for the LTE (dark line) and NLTE (light line)
models.  Models of $[{{\rm A}\over{\rm H}}]=-1$ (upper panel) and 
$[{{\rm A}\over{\rm H}}]=-4$ (lower panel).
\label{ppress}}
\end{figure}

Figs. \ref{ba24555_2} to \ref{sr24217_2} demonstrate the sensitivity of the results 
to the degrees of realism and completeness in the treatment of the NLTE problem by
comparing the NLTE$_{\rm Sr+Ba}$ and NLTE$_{\rm LTE}$ results to those of the NLTE$_{\rm Full}$
and LTE results.  
Note that we only carried out NLTE$_{\rm Sr+Ba}$ and NLTE$_{\rm LTE}$ modeling for values of 
$[{{\rm A}\over{\rm H}}]$ equal to
-1, -4, and -5, and in Figs. \ref{ba24555_2} to \ref{sr24217_2} we only show profiles for those
$[{{\rm A}\over{\rm H}}]$ values for which that particular line is unsaturated.  

\paragraph{}

We find that
in some cases (\ion{Ba}{2} $\lambda$ 5853.69, 6141.73 and 6496.91 for $[{{\rm A}\over{\rm H}}]=-1$)
the less realistic NLTE$_{\rm Sr+Ba}$ models yield line profiles that differ significantly
from those of the NLTE$_{\rm Full}$ model and are closer to the
LTE profiles than the NLTE$_{\rm Full}$ profiles are.  This indicates that these lines 
are indirectly affected by a significant amount by the non-local effect of overlapping 
transitions with other species that are also affected significantly by NLTE.  In the case of these
transitions of these species, the result of this ``multi-species non-locality'' is to 
enhance the deviation from the LTE results.  Therefore, spectrum synthesis calculations in 
which only the species of immediate interest are treated in NLTE (\ion{Sr}{0} and \ion{Ba}{0})
will underestimate NLTE corrections.  This result
points to the importance of a NLTE treatment for {\it all} species with significant
line opacity for the correct NLTE solution for {\it any one} species.  Particularly in
late type stars where the spectrum is blanketed by $\approx 10^7$ lines, there may
be any number of other species with transitions that overlap one or more of those of 
\ion{Sr}{0} or \ion{Ba}{0} whose equilibrium state thus determines the supply of 
photons that are available for absorption by the \ion{Sr}{0} or \ion{Ba}{0} transition
that is so overlapped.  (Moreover, the equilibrium state of those other species may in turn
be affected by yet {\it other} species that deviate from LTE whose transitions overlap
those of {\it those} species!)  The significance of this result bears careful consideration 
given the number of NLTE abundance determinations in the literature that are based on
LTE ``background'' line opacity for all species other than the one under investigation.
(Indeed, a way of understanding this result is that from the perspective of NLTE SE, there
is no such thing as ``background'' opacity; the set of NLTE SE equations non-locally couples
the states of {\it all} species.)  

\paragraph{}

However, we caution that the magnitude of this multi-species non-locality effect
for a particular line, or particular species, depends sensitively on the degree of 
wavelength overlap between lines of the species of interest that are important for
the excitation equilibrium, and strong lines of other
species treated in NLTE.  The degree of overlap and the magnitude of its influence, in turn, 
depends on the accuracy of 
the atomic $E$ level and oscillator strength ($gf$) data that is used to calculate the 
wavelengths and strengths, respectively, of the lines of overlapping species.   
The atomic data of \cite{kurucz94} has been produced originally for the purpose of
computing the broad-band solar flux level in LTE, for which only statistically correct
line opacity is important, and is known to contain erroneous level energies and
$gf$ values.  Therefore, the degree of the effect predicted here may be subject
to revision, either upward or downward, if more accurate atomic data were used for
all species in the NLTE calculation.  Due to the non-locality of this effect,
an exploration of the sensitivity of the effect to the perturbation of various 
$E$ levels of various species is beyond the scope of the present investigation.

\paragraph{}

In all cases
the profiles computed with NLTE$_{\rm LTE}$ modeling differ negligibly from those of
LTE modeling.  This indicates that the indirect effect of the NLTE opacity of other species
on the atmospheric structure is much less important than the direct effect of NLTE SE 
on the \ion{Ba}{0} and \ion{Sr}{0} equilibrium and the less direct effect of overlapping
transitions with other NLTE species.   This result is consistent with the minor effect on 
the $T$ structure of NLTE seen in Fig. \ref{models}.


\subsection{The role of collisional rates \label{res_coll}}

The greater realism of NLTE modeling comes at the expense of reliance on
a greater number of physical input data.  Among those that are
crucial to determining the NLTE SE equilibrium and that are among the least
well determined are the collisional cross-sections, $\sigma_{\rm ij}$, 
for $b-b$ and $b-f$ processes. 
The first $E$ level above the ground state of \ion{Ba}{2} has an excitation
energy of 0.604 ergs, which corresponds to a $T$ of ~4700 K for an ideal monatomic
gas.  Therefore, collisional $b-b$ transitions may be important for the excitation
equilibrium of \ion{Ba}{2}.
In our modeling we use approximate
$\sigma_{\rm ij}$ values for $e^-$ collisions calculated with the 
formula reproduced in \cite{lang99} for forbidden transitions
 for all $b-b$ 
transitions and of \citet{drawin61} for $b-f$
transitions.  Note that, for consistency of treatment among $b-b$ rates, and to
reduce reliance of uncertain $gf$ values, we 
do {\it not} use separate formulae for 
pairs of levels connected by permitted and forbidden transitions.
Adding to the uncertainty in collisional rates is the possible
role of collision partners other than electrons; MGB99 found that
collisions with H atoms play a detectable role in the
NLTE equilibrium state of \ion{Ba}{2} in main sequence and turn-off stars.  

\paragraph{}

To accommodate both of these considerations we perform a perturbation analysis by varying
{\it all} the $b-b$ and $b-f$ $\sigma_{\rm ij}$ values for \ion{Ba}{2} by a factor of
0.1 and 10.  According to Fig. \ref{ppress}, \ion{Ba}{1} is at most $10^{-3}$ times
as abundant as \ion{Ba}{2} and, thus, \ion{Ba}{1} rates are not expected to affect the
\ion{Ba}{2} equilibrium.  Moreover, there are no \ion{Ba}{1} lines that are useful for
abundance analysis.  Therefore, we leave the \ion{Ba}{1} rates unperturbed and 
vary separately both the $b-b$ and $b-f$ $\sigma_{\rm ij}$ values for \ion{Ba}{2} by a factor of
0.1 and 10.  This approach assumes that any errors in $\sigma_{\rm ij}$ values for either
$b-b$ or $b-f$ processes are systematic among all transitions rather than random.  However,
a transition-by-transition perturbation analysis, even if limited to critical transitions,
is beyond the scope of this investigation. 
The line opacity of \ion{Ba}{0} is not expected to affect 
the equilibrium atmospheric structure, so, with the $\sigma_{\rm ij}$ values at their
perturbed values, we re-converge the the \ion{Ba}{0} NLTE SE and recompute the 
spectrum using the NLTE$_{\rm Full}$ model structure.  

\paragraph{}

\begin{figure}
\plotone{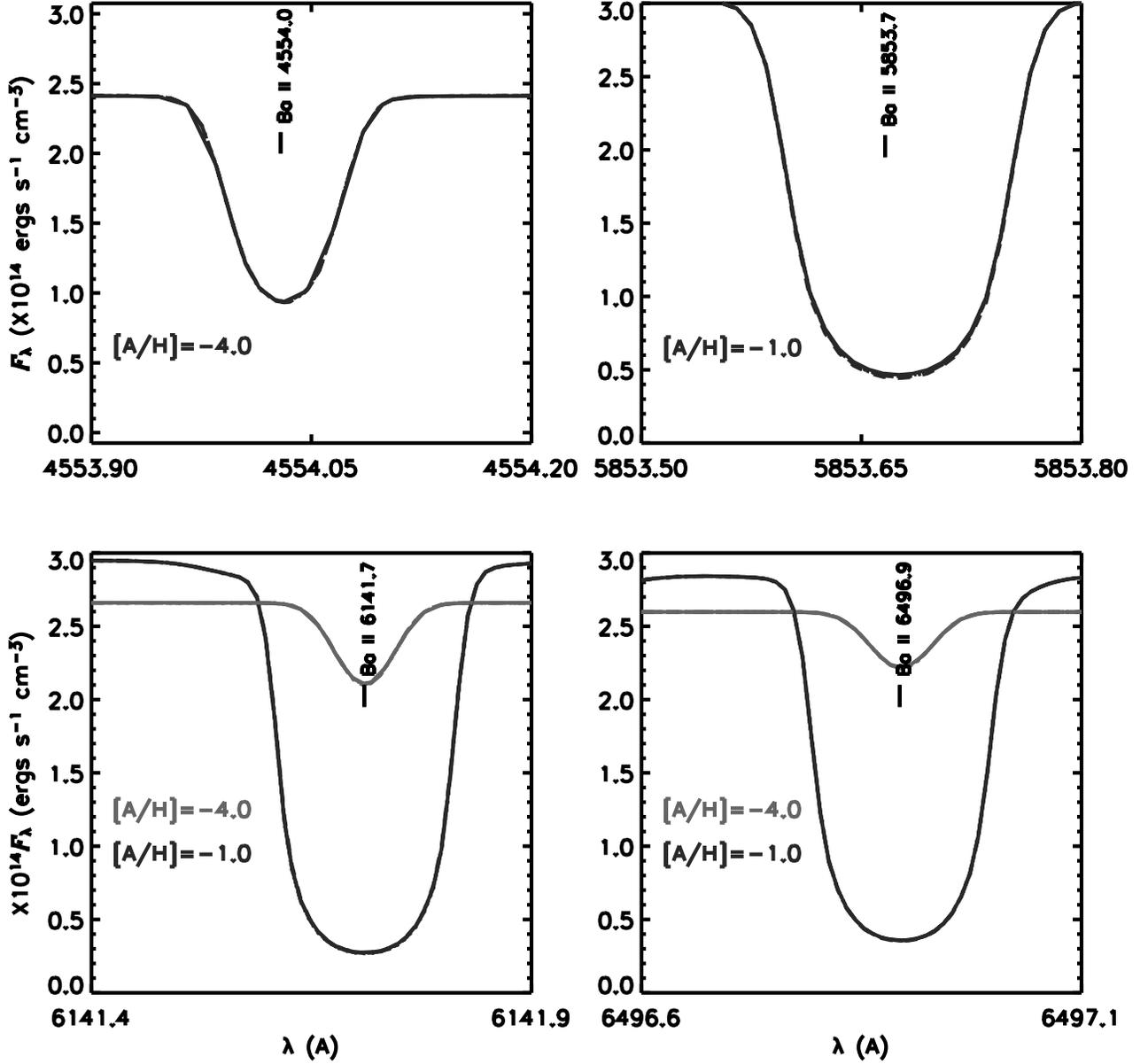}
\caption{Selected \ion{Ba}{2} lines computed with the NLTE$_{\rm Full}$ model
with the rates from standard analytic approximations (see text) (solid line),
with $b-b$ rates decreased (dotted line) and increased (dashed line) by a 
factor of 10, and with $b-f$ rates decreased (dot-dashed line) and increased 
(dot-dot-dot-dashed line) by a factor of 10.  Note that the perturbation 
results are barely visible because they differ negligibly from the result 
with standard formulae.
\label{ba24555_rate}}
\end{figure}

Fig. \ref{ba24555_rate} shows the computed \ion{Ba}{2} lines for the NLTE$_{\rm Full}$ model
with various values for the $\sigma_{\rm ij}$ collisional cross-sections.  
We find that perturbation of both the $b-b$ and the $b-f$ rates by a factor of 100 
produces both departure coefficients and line 
profiles that differ negligibly from the those calculated with the standard 
approximations.  This indicates that the excitation and ionization equilibria 
populations of 
\ion{Ba}{2} is robust.  In the case of the ionization equilibrium, the result is
expected given that \ion{Ba}{2} is the dominantly populated
ionization stage.  

\subsection{Hyperfine splitting of the \ion{Ba}{0} $\lambda 4554.03$ line}

The importance of hyper-fine splitting (HFS) of the \ion{Ba}{0} lines,
especially the $\lambda 4554.03$ line, has been
well documented by \citet{mcwilliam_a98}, \citet{mcwilliam_pss95}, \citet{mcwilliam_r94},
and references therein,
and these authors to various degrees present the formulae and data necessary for calculating 
the HFS components.  HFS has the effect of de-saturating lines and thus significantly alters the
inferred abundance from lines that are barely saturated and not yet heavily damped. 
For lines on the linear part of the COG, HFS changes the line profile, but not the value of
$W_\lambda$.
Because the hyper-fine sub-levels of a splitted level are relatively close to each other
in $E$, they are expected to be strongly coupled by collisional transitions, despite such
transitions being forbidden.  Therefore, all such sub-levels should have the same $b_{\rm i}$ 
value as the unsplitted level, and we thus expect that the effect of HFS on the profile of 
a line computed in NLTE will be the same as that for a line computed in LTE.
However, for the sake of completeness, and because the HFS line profile is of interest,
we investigate the effect of HFS in NLTE.

\paragraph{}

We have modified our \ion{Ba}{2} atomic model to account for HFS in the $\lambda 4554.03$ resonance
line where the effect is known to be especially important.  The structure of the {\tt PHOENIX} code is such
that it is more straightforward to represent the splitted sub-levels directly in the atomic model (the way
nature does it) rather than by mimicing the effect of splitting by modifying the line profile function, 
$\phi_\nu$, as is suggested by, for example, \cite{gray_d92}.  We deduced the $E$ level splitting and the
component $\log gf$ values from the component $\lambda$ and relative $gf$ values given
in Table 1 of \citet{mcwilliam_a98}.  In stars, \ion{Ba}{0} may be significantly
abundant over the range of atomic masses, $A$, from 134 to 138.  The component $gf$ values 
are weighted by the relative abundance of the isotopes of odd $A$ value because only they
have HFS.  Because we are modeling metal-poor red giants, we 
follow \citet{mcwilliam_a98} and \citet{sneden_mpcba96}, and adopt a pure $r$-process 
isotopic mix of 0:40:0:32:28 for isotopes of A value of 134, 135, 136, 137, and 138, respectively,
which amounts to an odd to even mix of 72:28.  

\paragraph{}

\begin{figure}
\plotone{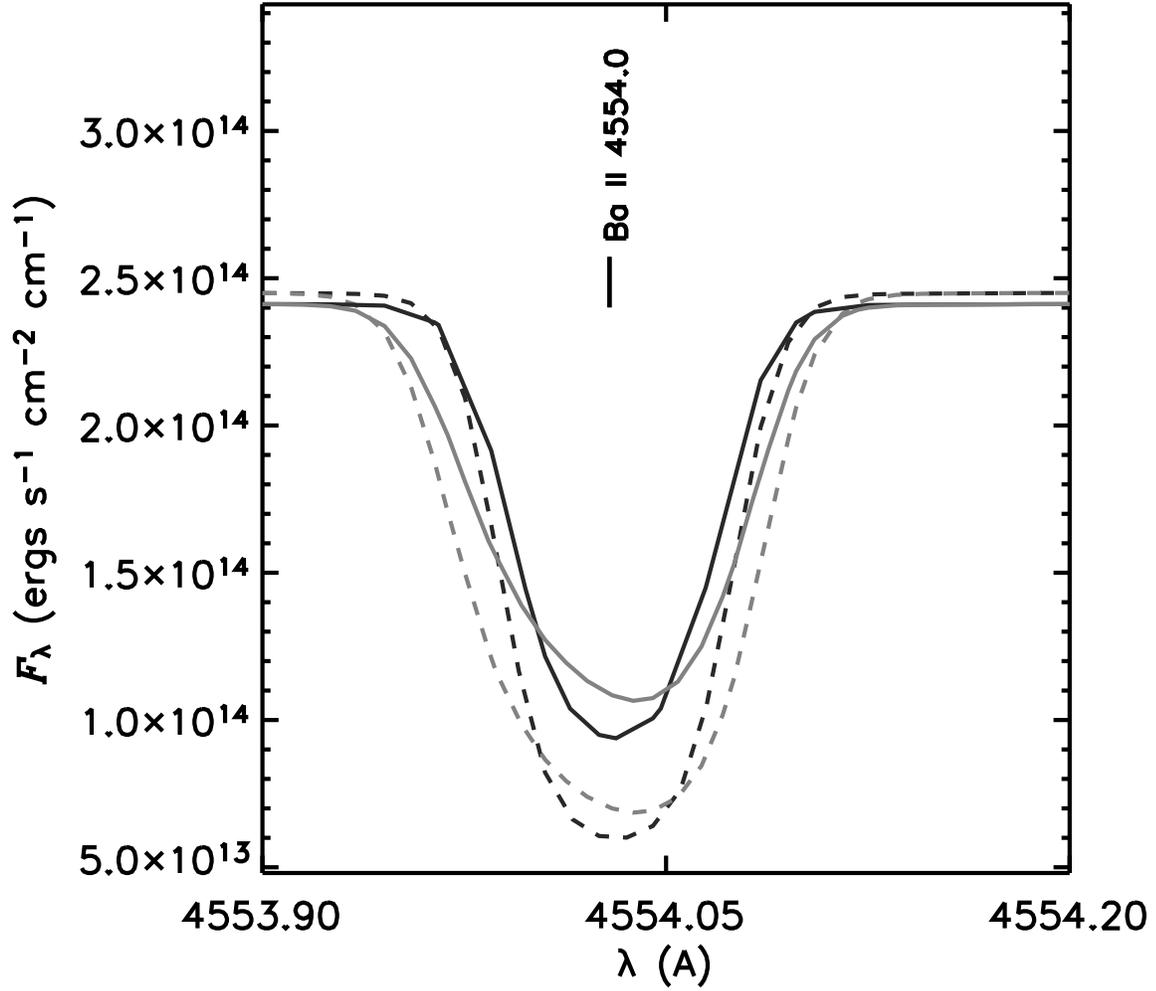}
\caption{As for Fig. \ref{ba24555} for the model of $[{{\rm A}\over{\rm H}}]=-4$ 
with the addition of the model with hyper-fine splitting (HFS) (gray lines). 
\label{ba24555_hfs}}
\end{figure}

Fig. \ref{ba24555_hfs} shows the $\lambda 4554.03$ line computed with the NLTE$_{\rm Full}$ and 
LTE models with and without HFS in the treatment.  We investigate the model of 
$[{{\rm A}\over{\rm H}}]$ value equal to -4 for which the line is clearly on the linear part of
the COG.  As expected, the effect of HFS is to broaden the line while preserving its strength.
We note that HFS also makes the line profile asymmetric as a result of the asymmetry of the 
HFS components.  Also as expected, the effect of NLTE on the profile computed with HFS is 
the same as that for the profile computed without HFS; the extent of the NLTE weakening of 
the line is independent of the $E$ level splitting.

\section{Conclusions \label{sec_con}}

From comparison of LTE models with the most realistic NLTE$_{\rm Full}$ models,
we find that for all lines that are useful \ion{Sr}{0} and \ion{Ba}{0} abundance diagnostics
the magnitude and sense of the computed NLTE effect on the predicted line strength 
is metallicity dependent, with stars of $[{{\rm A}\over{\rm H}}]= -1$ to
$-2$ showing NLTE strengthening of the lines and those of $[{{\rm A}\over{\rm H}}]= -4$ to $-5$
showing NLTE weakening.  These results are qualitatively similar to those of MGB99 for
main sequence and turn-off stars and point to 
the need to account for NLTE effects
when deriving spectroscopic \ion{Sr}{0} and\ion{Ba}{0} abundances. 

\paragraph{\ion{Ba}{0}: }
From Table \ref{t3}, we see that among the most affected lines is 
the \ion{Ba}{0} $\lambda 4554.03$ (resonance) for $[{{\rm A}\over{\rm H}}]= -4$, for which 
$\log W_\lambda$(NLTE$_{\rm Full}$)-$\log W_\lambda$(LTE)=-0.14 (neglecting HFS), leading to 
a NLTE abundance correction of $\epsilon_{\rm NLTE}-\epsilon_{\rm LTE}=+0.14$ (because the line
is on the linear part of the COG).  For 
\ion{Ba}{0} $\lambda 5853.69$ in models of $[{{\rm A}\over{\rm H}}]= -1$, 
$\log W_\lambda$(NLTE$_{\rm Full}$)-$\log W_\lambda$(LTE)=+0.04, leading to 
$\epsilon_{\rm NLTE}-\epsilon_{\rm LTE}=-0.04$.  

\paragraph{\ion{Sr}{0}: }
Similar effects, though less pronounced, are found for the two \ion{Sr}{0} lines.  We note that these
lines are so strong that they are only on the linear part of the COG for the most metal poor of our
models ($[{{\rm A}\over{\rm H}}]=-4$ to $-5$).

\paragraph{}

 A less well studied NLTE effect is multi-species non-locality that arises from overlapping
transitions that link the equilibrium state of one species with those of other species whose
equilibrium also departs from LTE.  By including most light metals up to and including much of the \ion{Fe}{0}
group in NLTE (NLTE$_{\rm Full}$ models), we find that for some lines such overlap enhances NLTE 
abundance corrections non-negligibly, and should, in principle, be accounted for for accurate abundance 
determination.  From Table \ref{t3} the largest such effect is for \ion{Ba}{0} $\lambda 5853.69$
for $[{{\rm A}\over{\rm H}}]= -1$ for which $\log W_\lambda$(NLTE$_{\rm Full}$)-$\log W_\lambda$(Sr+Ba)=+0.03.
However, we reiterate that the magnitude of this effect is dependent on the accuracy of the atomic
data of the overlapping NLTE species.  Given the notorious innacuracy of such data, caution is
required on this point.
Another indirect NLTE effect is that of the opacity of other species that depart from LTE 
that are important for radiative equilibrium of the atmosphere.  We find that the NLTE
$T$ structure is close enough to that of LTE in the line forming region that 
the effect on the computed \ion{Sr}{0} and \ion{Ba}{0} line profiles is negligible.  We conclude
that NLTE spectrum formation calculations with LTE atmospheric models will suffice for 
red giants.  Freedom from the necessity to calculate large grids of NLTE atmospheric models
for large NLTE abundance surveys will expedite such work.

\paragraph{}

We find that the computed NLTE line strengths differ negligibly if
collisional $b-b$ and $b-f$ rates are an order of magnitude larger or smaller than 
those calculated with standard analytic formulae.  Because collisional 
cross-sections are notoriously uncertain, the robustness against perturbation in this
case provides some assurance that the results are not artifact of erroneous
atomic data.

\paragraph{}

For completeness, we have also studies the effect of NLTE on the 
\ion{Ba}{0} $\lambda 4554.03$ resonance lines when HFS is taken into account.
HFS for \ion{Ba}{0} has been extensively studied elsewhere and we reproduce 
the well-known result that HFS substantially widens the line and increases
the residual flux at line center.  
Of note here is that, as expected from the close collisional coupling among the
splitted sub-levels, the relative effect of NLTE is the same with and without HFS.



\acknowledgments

CIS gratefully acknowledges funding from the Natural Sciences and Engineering 
Research Council of Canada (grant no. 264515-03), the Canada Foundation for
Innovation (project no. 9272), and Saint Mary's University.  Some of these
calculations were performed on the pluto cluster of the Institute for Computational
Astrophysics, funded by
the Canada Foundation for Innovation. 
This work was supported in part by NSF grants AST-9720704 and  
AST-0086246, NASA
grants NAG5-8425, NAG5-9222, as well as NASA/JPL grant 961582 to the  
University
of Georgia.  This work was supported in part by the P\^ole Scientifique  
de
Mod\'elisation Num\'erique at ENS-Lyon.  Some of the calculations  
presented in
this paper were performed on the IBM pSeries 690 of the Norddeutscher  
Verbund
f\"ur Hoch- und H\"ochstleistungsrechnen (HLRN), on the IBM SP  
``seaborg'' of
the NERSC, with support from the DoE.  We thank all these institutions 
for a generous allocation of
computer time.





\clearpage

\clearpage

\clearpage

\clearpage

\clearpage

\clearpage

\clearpage

\clearpage

\clearpage

\clearpage

\clearpage 

\begin{table}
\caption{Degrees of modeling realism.}
\label{t1}
\begin{tabular}{llll}
\tableline
Model            & Species in            & Atmospheric \\
                 & NLTE SE               & structure   \\
\tableline
NLTE${\rm Full}$ & Full set              & NLTE        \\
NLTE${\rm Sr+Ba}$& \ion{Sr}{0}, \ion{Ba}{0}& NLTE        \\
NLTE${\rm LTE}$  & None                  & NLTE        \\
LTE              & None                  & LTE         \\
\\
\tableline
\end{tabular}
\end{table}

\clearpage 

\begin{table}
\caption{\ion{Sr}{0} and \ion{Ba}{0} \ion{}{1} and \ion{}{2} atomic models.}
\label{t2}
\begin{tabular}{lllll}
\tableline
Species  & No. $E$-levels & No. $b-b$ transitions & $\chi_{\rm I}$ (eV) \\
\tableline
Sr\,{\sc I} & 52  & 74 & 5.70 \\
Sr\,{\sc II} & 32 &  90 & 11.03\\
Ba\,{\sc I} & 76 &  114 & 5.21 \\
Ba\,{\sc II} & 51 &  121  & 10.00\\
\tableline
\end{tabular}
\end{table}

\clearpage

\begin{table}
\caption{Selected atomic data for \ion{Sr}{2} and \ion{Ba}{2} lines of diagnostic
utility.  Data are from \citet{kurucz94}, except for level designations, which are from NIST.}
\label{t2_5}
\begin{tabular}{llllll}
\tableline
$\lambda$ \AA~ (Air) & $A_{\rm ji}$ (s$^{-1}$) & $\chi_{\rm i}$ (cm$^{-1}$) & $\chi_{\rm j}$ & Lower level & Upper level\\ 
\tableline
\ion{Ba}{2}\\
\tableline
4554.033             & 1.11e+08     & 0.000       & 21952.404 & [\ion{Xe}{0}]$6s ^2S_{1/2}$ & [\ion{Xe}{0}]$6p ^2P^o_{3/2}$   \\ 
5853.675             & 6.00e+06     & 4873.850    & 21952.404 & [\ion{Xe}{0}]$5d ^2D_{3/2}$ & [\ion{Xe}{0}]$6p ^2P^o_{3/2}$   \\
6141.713             & 4.12e+07     & 5674.824    & 21952.404 & [\ion{Xe}{0}]$5d ^2D_{5/2}$ & [\ion{Xe}{0}]$6p ^2P^o_{3/2}$   \\
6496.898             & 3.10e+07     & 4873.852    & 20261.561 & [\ion{Xe}{0}]$5d ^2D_{3/2}$ & [\ion{Xe}{0}]$6p ^2P^o_{1/2}$   \\
\tableline
\ion{Sr}{2}\\
\tableline
4077.71              & 1.42e+08     & 0.000       & 24516.650 & [\ion{Kr}{0}]$5s ^2S_{1/2}$ & [\ion{Kr}{0}]$5p ^2P^o_{3/2}$  \\
4215.52              & 1.27e+08     & 0.000       & 23715.190 & [\ion{Kr}{0}]$5s ^2S_{1/2}$ & [\ion{Kr}{0}]$5p ^2P^o_{1/2}$ \\
\tableline
\end{tabular}
\end{table}

\clearpage

\begin{table}
\caption{Equivalent widths, $W_\lambda$, for lines used in the abundance analysis of JFBCGS04 under
various modeling assumptions, shown for select $[{{\rm A}\over{\rm H}}]$ values.  For unsaturated
lines, the inferred abundance in $[{{\rm A}\over{\rm H}}]$ units is proportional to $\log W_\lambda$.}
\label{t3}
\begin{tabular}{lllllll}
\tableline
                   &       & $W_\lambda$ (m\AA) \\
Line                      & $[{{\rm A}\over{\rm H}}]$ & NLTE$_{\rm Full}$ & NLTE$_{\rm Sr+Ba}$ & NLTE$_{\rm LTE}$ & LTE \\
\tableline
Ba\,{\sc II} $\lambda 4554.03$ & -4                        & 48.4            & 49.2             & 66.6           & 66.3 \\
Ba\,{\sc II} $\lambda 5853.69$ & -1                        & 132.2           & 124.3            & 121.8          & 120.9 \\
Ba\,{\sc II} $\lambda 6141.73$ & -1                        & 182.7           & 177.5            & 172.7          & 170.9 \\
Ba\,{\sc II} $\lambda 6141.73$ & -4                        & 16.1            & 17.5             & 17.6           & 17.4 \\
Ba\,{\sc II} $\lambda 6496.91$ & -1                        & 179.3           & 171.3            & 167.7          & 165.9 \\ 
Ba\,{\sc II} $\lambda 6496.91$ & -4                        & 11.8            & 12.7             & 12.8           & 12.6 \\ 
\\
Sr\,{\sc II} $\lambda 4077.71$ & -5                        & 47.2            & 49.0             & 55.0           & 55.0 \\
Sr\,{\sc II} $\lambda 4215.52$ & -5                        & 32.6            & 34.1             & 39.0           & 39.0 \\
\tableline
\end{tabular}
\end{table}

\end{document}